*Article*

# From an Entropic Measure of Time to Laws of Motion

**Leonid M. Martyushev[1,2,*], Evgenii V. Shaiapin[1]**

1   Ural Federal University, 19 Mira St., Ekaterinburg, 620002 Russia; shayapin@mail.ru
2   Institute of Industrial Ecology, Russian Academy of Sciences, 20 S. Kovalevskaya St., Ekaterinburg, 620219 Russia
*   Correspondence: LeonidMartyushev@gmail.com; Tel.: +7-922-22-77425

**Abstract:** A hypothesis proposed in the paper [Entropy 2017, 19, 345] on the deductive formulation of a physical theory based on explicitly- and universally-introduced basic concepts is further developed. An entropic measure of time with a number of properties leading to an analog of the Galileo–Einstein relativity principle is considered. Using this measure and a simple model, a kinematic law which relates time to the size and number of particles of a system is obtained. Corollaries of this law are examined. In particular, accelerated growth of the system size is obtained, whereas in systems with constant size decrease in the number of particles is observed. An interesting corollary is the emergence of repulsive and attractive forces inversely proportional to the square of the system size for relatively dense systems and constant for systems with sufficiently low density.

## 1. Introduction

By studying certain properties of the surrounding world, a physicist would usually formulate some tentative theory. Such theories are usually composed of three parts: (1) basic (initial) concepts, (2) laws (or principles), and (3) a number of assumptions and idealizations (model). In the case of mechanics[1], the first part may consist of concepts of time, space, mass, etc.; the second may be represented by the Newton laws, including the law of gravity and the principle of least action. In mechanics there are many assumptions and idealizations. For example, these include point mass, free body, closed system, perfectly elastic collision, Euclidean (as well as isotropic, homogeneous) space, etc. When corollaries of the tentative theory are analyzed, especially when compared with experiment, a number of problems arise which, in their turn, lead to refinements and modifications of the tentative theory (see Fig.1a). Thanks to such refinements, contradictions between components forming three parts of the theory are eliminated or smoothed out. It should be noted that experimental data are interpreted within the framework of the tentative theory in its current, provisional form. After a number of iterations, the theory looks more or less complete (to the majority of scientists) for a rather long period until some new "unexpected" problem comes up. This results in new and sometimes substantial adjustments in the tentative theory causing the so-called scientific revolution. This is a well-known process described by historians and philosophers of science long ago (specifically by K. Popper, T. Kuhn, and I. Lakatos) (Fig.1a). In the history of physics and technology, this method proves to be very effective and productive. However, such a development of theory has methodological disadvantages. Let us list them.

1. Basic concepts (the "language" of a theory) are usually based neither on measurements nor on logical operations. They are introduced by definition (axiomatically), based on conscious and/or subconscious experience gained through interactions with the surrounding world (observations and experiments). The best example of this is the introduction of time in mechanics (see, e.g., the well-known and respected textbook on theoretical physics [1]). The conventional description fails to give any idea how time should be measured or introduced (without being based on laws of mechanics obtained only on the assumption that time exists and has some properties). Why can time be considered homogeneous, regular, expressed in real numbers, etc.? These questions are not new and were raised many times, e.g. by E. Mach, H. Poincaré and P. Bridgman.

---

[1] In what follows, we shall base our reasoning on mechanics as the most developed and fundamental branch of physics.

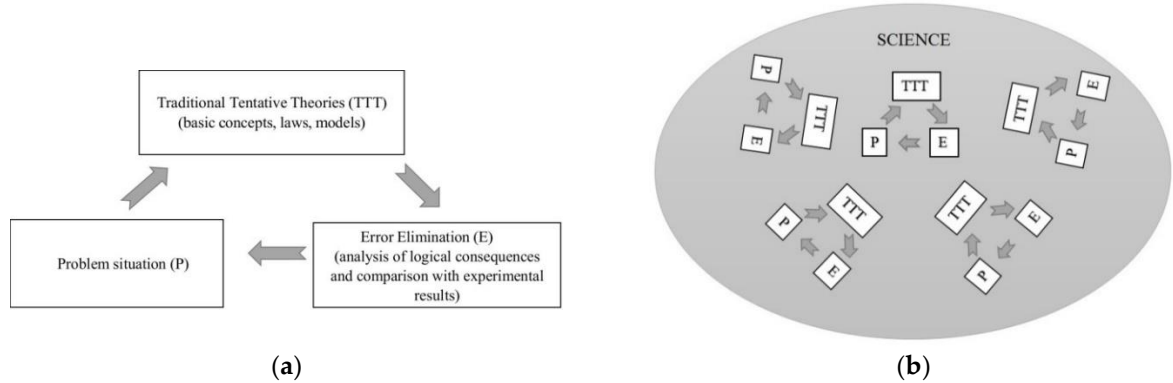

**Figure 1.** The evolution of tentative theory (a) and science (b). The traditional method.

2. Theories as a whole (basic concepts and laws/principles as well as assumptions and idealizations) adapt to experimental data. In the case of mechanics this led to assertions, for example, that if all coordinates and velocities are specified, it allows fully determining the state of a mechanical system and predicting its motion, or for instance, that gravitational force is exactly inversely proportional to the square of the distance between bodies [1]. All three parts of a theory (basic concepts, laws, and models) effectively have one origin and validation tool: observation and experiment. In the light of the above, dividing a theory into three parts is in many respects a very approximate procedure. In essence, these parts are identical, they form a *single interrelated system* that evolves historically, following the diagram in Fig.1a, to describe some natural phenomena. Due to the absence of strong logical connections between the parts of a theory, its iterative adaptations to experimental data enable to change any of its elements rather freely (a typical example here is the well-known "transformation" of classical mechanics in quantum or relativistic cases when problems with experimental results were found). Such an optimization is accompanied by an unlimited number of degrees of freedom. As a result, such theories can no longer be trusted as unique and fundamental. In spite of this conclusion, the history of science shows that basic concepts and laws (principles) are often assigned meanings that essentially go beyond the framework of their original tentative theories and are regarded as universally valid. For example, the laws of classical thermodynamics are freely applied to cosmology. Obviously, given the origin of a tentative theory, such an "extrapolation" can prove to be either very bold and fruitful or risky and erroneous.

3. Different tentative theories exploring different, sometimes unrelated areas of science (not necessarily only natural science) can have similar basic concepts. However, every tentative theory develops *independently* (see Fig.1b). Iterations and optimizations in individual theories can lead to basic concepts acquiring increasingly different properties. As a result, science is divided into weakly interacting parts, scientific knowledge loses its universality, the gap of understanding between scientists of various specializations widens. Undoubtedly, all this has a negative effect on the scientific progress as a whole. For example, the concept of time in science now has a different meaning not only for biologists, psychologists, and historians but also for physicists in different fields: thermodynamics, classical and quantum mechanics, or cosmology.

A possible methodological way out of this situation was previously suggested in [2]. It is schematically shown in Fig.2a. Let us comment on this diagram. Basic concepts (for example, time) should be formulated in such a way that they can be defined (measured, calculated) in a uniform manner for every branch of the modern science (liberal arts included). Then, by studying some phenomenon of the surrounding world, a model (a set of axioms, idealizations, and assumptions) should be made. Thereafter, using the introduced basic concepts and model, laws (corollaries) should be ***deductively*** obtained and then compared with experiment and analyzed. In case contradictions or problems are found, the model should be adjusted with the cycle repeating. Further, the basic concepts should be treated in an extremely conservative way. These concepts concern and bind together the entire scientific framework and its multiple theories thus forming the basis of our scientific worldview (Fig.2b). So, in comparison with the old method of theory formulation (Fig.1), the new one has the following main features:

1) Basic concepts are introduced explicitly, operationally (constructively) and universally. On these specific grounds, the concepts may be called basic.

2) Laws are a logical corollary of the deductive development of basic concepts with the help of models. So, the strict hierarchy is established: *basic concept + model → law.*

3) During the analysis and comparison with experiment, only elements of the model are optimized. As a result, there is a drastic reduction in the number of degrees of freedom when bridging the gap between theoretical predictions and experimental data.

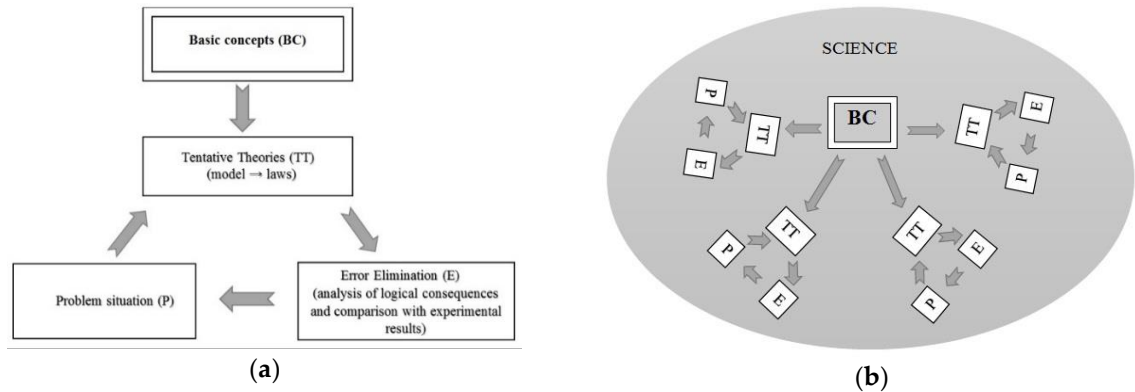

**Figure 2.** The evolution of tentative theory (a) and science (b). The proposed method.

This method is unusual for physics. However, similar ideas were expressed before. Notably, E. Milne established a number of basic concepts and a model of the expanding Universe in order to logically derive a number of conclusions in mechanics and electromagnetism which he later compared with traditionally-established laws [3]. This is a very interesting, original and comprehensive study. In his approach, time (clock) is a primary basic concept which is related to the properties of light rays emitted and received by various observers. In Milne's theory, spatial interval is a secondary concept. In our opinion, the lack of universality in the introduced concept of time is a significant drawback of this method. Indeed, the concept is introduced using light rays the physical properties of which (including constant velocity in vacuum) determine the properties of time. Such "time" can hardly be introduced and used in other scientific disciplines.

In the paper [2] (developing ideas mentioned in [4,5]), a measure of time related to the concept of informational entropy was introduced. The essential properties of this measure are its universality (i.e., applicability to both natural sciences and liberal arts) and its constructive potential. The latter means that on the basis of this measure, one could deduce dynamic laws, if required. In comparison with the traditional approach, such a method of developing a physical theory (e.g., mechanics), is rather new and promising. As a consequence, the purpose of this paper is to obtain, based on such time measure (and using a simple model), a number of laws of mechanics which could be subsequently compared with experiment.

## 2. Measure of time and model

Let us assume according to [2] that time $\tau$, up to a dimensionless[2] constant, is equal to a logarithm of the number of possible microstates ($\Gamma$) representing some state (macrostate) at a given moment of observation (or entropy $S$)[3].

$$\tau = \varepsilon \ln \Gamma \text{ or } \tau = \varepsilon S, \qquad (1)$$

where $\varepsilon$ is some number. Let us also accept two postulates with regard to time [2]: 1) Time exists (is defined) only if changes in the observer–observed system occur; 2) its value may increase only. These postulates restrict potential changes in the system (specifically, they rule out changes that result in $\Gamma$ decreasing or being constant).

---

[2] All the quantities assumed herein, particularly time and size, are dimensionless. This does not impose any fundamental restrictions but only simplifies notation.
[3] In the considered model entropy turns out to be a familiar configurational (Boltzmann's) entropy.

Let us discuss a crucial issue arising when complex systems (i.e., systems consisting of multiple coexisting subsystems) are considered using the introduced measure of time. An individual time scale can be assigned to each subsystem using entropy, and this poses a question: are these time scales uniform or not in relation to each other? In other words, are elementary intervals of the change of time $d\tau$ for two different scales nonlinearly connected? The answer is no because, as is known from the maximum entropy production principle [6,7], coexisting subsystems have the same change of entropy (entropy productions). According to Eq.(1), the change of time $d\tau$ is directly related to the change of entropy, and as a result the time scales are linearly connected. As a consequence, for such subsystems, laws derived from these time measures using the same model are similar. In this regard, it can be stated that laws for different coexisting subsystems are the same. This statement can be considered as an analog of the Galileo–Einstein relativity principle.

Let us define the measure of length in a traditional way using some accepted standard (e.g., perfectly rigid rod). Using this measure of length, we shall represent the available space as the number $G$ of identical cells. The number of cells $G$ thus represents the size of the system under consideration.

Let us consider the following model. There are two types of identical particles $b$ and $f$ in the space of size $G$ the number of which are $\alpha N$ and $(1-\alpha)N$, respectively (where $N$ is the total number of particles in the system and $\alpha$ is the fraction of particles $b$, $\alpha \in [0;1]$). Particles $b$ and $f$ differ in terms of their possible distribution in the cells of the space. Let us assume that one cell of the space can contain any number of particles $b$ (up to $N$ if $\alpha = 1$) and not more than one particle $f$. Particles with such properties resemble the well-known particles: bosons and fermions. The principal difference is that in the case of bosons and fermions, cells are primarily associated with energy rather than spatial states, and the number of cells is usually considered constant rather than variable as in our case.

By the number of microstates we shall mean the number of possible distributions of particles $b$ and $f$ among $G$ cells. A macrostate shall mean a system with some $G$, $\alpha$, and $N$ values.

Thus, our model contains four quantities ($\tau$, $G$, $\alpha$, and $N$). Of these, $\tau$ and $G$ can be regarded as basic, whilst the other two in the proposed model are auxiliary. Let us make three assumptions. (1) Particles $b$ and $f$ are distributed in cells independently. (2) Time is defined by a specific (per the number of particles) entropy, i.e. $\varepsilon = 1/N$. (3) The fraction of particles $\alpha$ in the system shall be such that the system's specific entropy reaches its maximum value.

### 3. Corollaries of the measure of time and the model: a law

It is well known [8] that the number ($\Gamma_1$) of possible microstates for particles $b$ corresponding to a given macrostate is:

$$\Gamma_1 = \frac{(\alpha N + G - 1)!}{(\alpha N)!(G-1)!}, \tag{2}$$

the number ($\Gamma_2$) of possible microstates for particles $f$ is:

$$\Gamma_2 = \frac{G!}{((1-\alpha)N)!(G-(1-\alpha)N)!}. \tag{3}$$

Using the first assumption of the model, we have:

$$S = \ln \Gamma_1 + \ln \Gamma_2, \tag{4}$$

for the specific (per $N$) entropy $s$, based on the Stirling's formula and assuming that $G \gg 1$, $N \gg 1$ and $G \gg (1-\alpha)N$, it can be obtained that:

$$s = \frac{S}{N} = \alpha\left(1 + \frac{G}{\alpha N}\right)\ln\left(1 + \frac{G}{\alpha N}\right) - (1-\alpha)\left(\frac{G}{(1-\alpha)N} - 1\right)\ln\left(\frac{G}{(1-\alpha)N} - 1\right) + \frac{G}{N}\ln\left(\frac{\alpha}{(1-\alpha)}\right). \tag{5}$$

Using the third assumption of the model, let us find an optimal value of $\alpha$ for which the system's specific entropy would be at its maximum. Differentiating (5) with respect to $\alpha$ and setting it equal to zero, we obtain:

$$\frac{\partial s}{\partial \alpha} = \ln\left(\frac{\alpha N + G}{\alpha N}\right) - \ln\left(\frac{G - N + \alpha N}{(1-\alpha)N}\right), \tag{6}$$

$$2N\alpha^2 + 2(G-N)\alpha - G = 0. \tag{7}$$

Subject to conditions $N > 0$, $G > 0$, $\alpha \in (0;1)$, we get the unique solution of the last equation:

$$\alpha_m = \frac{N - G + \sqrt{N^2 + G^2}}{2N}. \tag{8}$$

Using (5) and (8), and applying the second assumption of the model, we get:

$$\tau = \left(\frac{G}{N} - \frac{1}{2}\right)\ln\left(N - G + \sqrt{N^2 + G^2}\right) - \left(\frac{G}{N} + \frac{1}{2}\right)\ln\left(N + G - \sqrt{N^2 + G^2}\right) + \\ + \frac{1}{2}\ln\left(N + G + \sqrt{N^2 + G^2}\right) + \frac{1}{2}\ln\left(G - N + \sqrt{N^2 + G^2}\right) \tag{9}$$

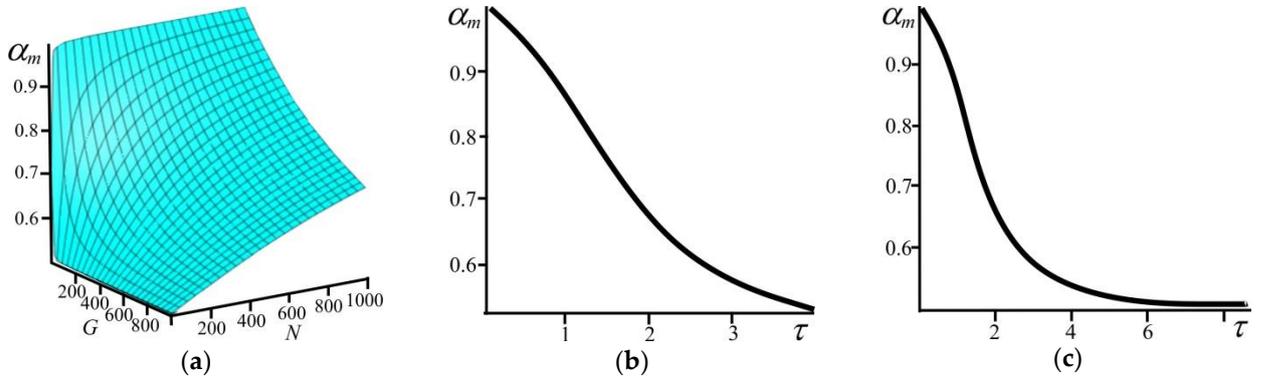

**Figure 3.** The fraction of particles $\alpha_m$ as a function of the system size $G$ and the number of particles $N$ (a). The fraction of particles $\alpha_m$ as a function of the time $\tau$ with N=$10^3$ (b) and with G=$10^3$ (c).

Expression (law) (9) interrelates all the model quantities. We shall call it the kinematic relationship. It is valid for times larger than the time of maximization of the specific entropy due to redistribution of particles $b$ and $f$ in cells. Using (8) and (9), we can plot variation of fractions of particles $b$ and $f$ with time (Fig.3). It can be seen in Fig.3 that with the passage of time the fraction of particles $b$ decreases from 1 in the superdense state ($N/G \to \infty$) to 0.5 in the low dense state ($N/G \to 0$).

Fig.4, based on Eq.(9), shows how the size of the system and the number of particles depend on time. It can be concluded from Fig.4 that for any values of $G$ and $N$ time is greater than zero.

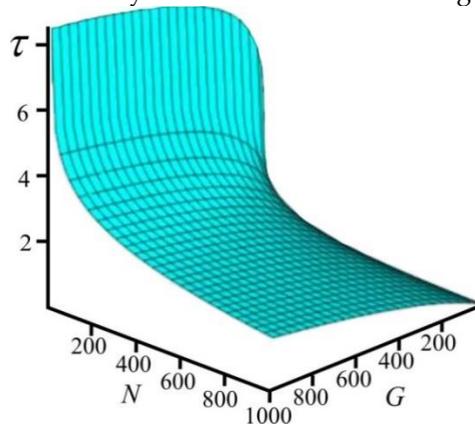

**Figure 4.** The relationship between time $\tau$, the system size $G$ and the number of particles $N$.

According to the second postulate, time in the system can only increase. This allows determining the restrictions on possible variations of the parameters. Indeed,

$$d\tau = \frac{\partial \tau}{\partial N}dN + \frac{\partial \tau}{\partial G}dG > 0. \tag{10}$$

According to (9):

$$\frac{\partial \tau}{\partial N} = -\frac{G}{N^2} \cdot \Omega(N,G), \tag{11}$$

$$\frac{\partial \tau}{\partial G} = \frac{1}{N} \cdot \Omega(N,G). \tag{12}$$

where $\Omega(N,G) \equiv \ln\left(\frac{N-G+\sqrt{N^2+G^2}}{N+G-\sqrt{N^2+G^2}}\right)$. We can show that $\Omega > 0$.

As a result, (10) is reduced to the form:

$$\frac{dG}{G} > \frac{dN}{N}. \tag{13}$$

So, in order for time to increase, a relative change of the system's size must be greater than a relative change of the number of particles in the system. In the particular case when the number of particles remains constant, according to (13) the system's size must inevitable increase, whereas when the system evolves maintaining its constant size, the number of particles must decrease at all times. Fig.4 illustrates exactly this behavior.

The kinematic relation Eq.(9) is rather complex. However, it can be considerably simplified for two extreme cases.

1. The superdense state. In the extreme case where $N/G \to \infty$, from Eq.(9) we obtain:

$$\tau \approx \frac{G}{N}\ln\left(\frac{2Ne^{1/2}}{G}\right), \tag{14}$$

or

$$G \approx -\frac{\tau \cdot N}{W_{-1}\left(-\frac{\tau}{2e^{1/2}}\right)}, \tag{15}$$

where $W_{-1}$ is the lower real-valued branch of the Lambert-W function (this branch was chosen assuming that for $N/G \to \infty$ time must be positive) [9].

As can be seen from Fig.5, for the superdense state $G/N$ grows with the increase of time in an approximately linear manner.

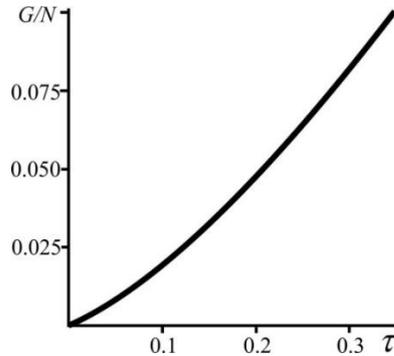

**Figure 5.** The dependence of $G/N$ on time $\tau$ for the superdense state ($N/G \to \infty$), based on Eq.(15).

2. The state of low density. In the extreme case where $N/G \to 0$, from Eq.(9) we obtain:

$$\tau \approx \ln\left(\frac{2G}{N}\right), \tag{16}$$

or

$$G \approx \frac{N}{2} \cdot e^{\tau}. \tag{17}$$

It can be seen that, in case of sufficiently low density, $G/N$ grows with time exponentially.

## 4. Corollaries from the derived law

Let us discuss some results directly following from the kinematic law derived above.
1. We define the rate of the system's size change as $v=(\partial G/\partial \tau)_N$. According to Eq.(12):

$$v = \left(\frac{\partial \tau}{\partial G}\right)_N^{-1} = N \cdot \Omega(N,G)^{-1}. \tag{18}$$

The rate of change of the size of the system as a function of its size, number of particles and time, as calculated on the basis of Eqs.(9) and (18), is shown in Fig.6. It can be seen from Fig.6a that $v$ grows with the increase of $G$ and $N$, and this change in $G$ occurs much faster. With time, $v$ increases too (Fig.6b).

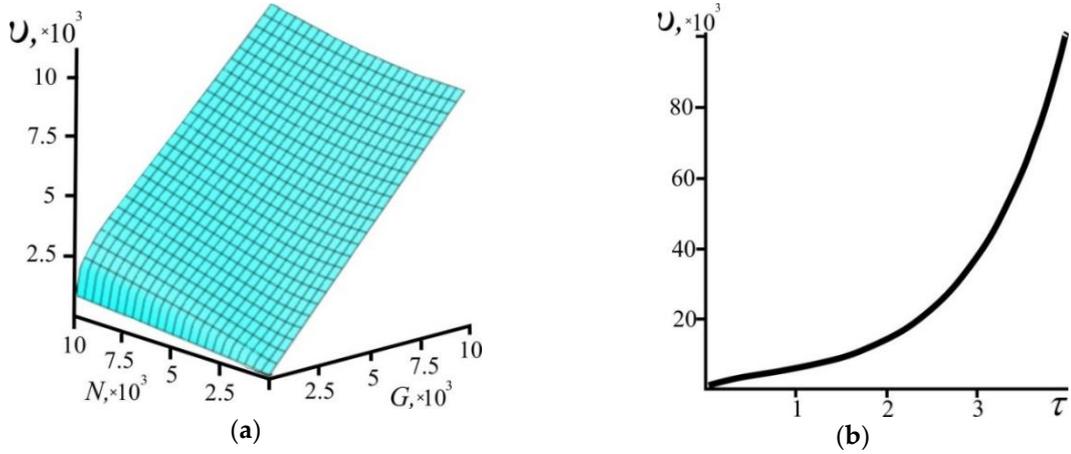

(a)      (b)

**Figure 6.** The rate of change of the system size $v$ as a function of the system size $G$ and the number of particles $N$ (a) and as a function of the time $\tau$ with $N=10^4$ (b).

For the two extreme cases, dependences of velocity on time can be obtained explicitly. For the superdense state, using (15) and the property of the Lambert function [9]:

$$W'(x) = \frac{W(x)}{x \cdot (1+W(x))}, \tag{19}$$

we obtain:

$$v = -\frac{N}{1+W_{-1}\left(-\dfrac{\tau}{2e^{1/2}}\right)}. \tag{20}$$

For the state of low density, according to (17), we have:

$$v = \frac{N}{2} \cdot e^{\tau}, \tag{21}$$

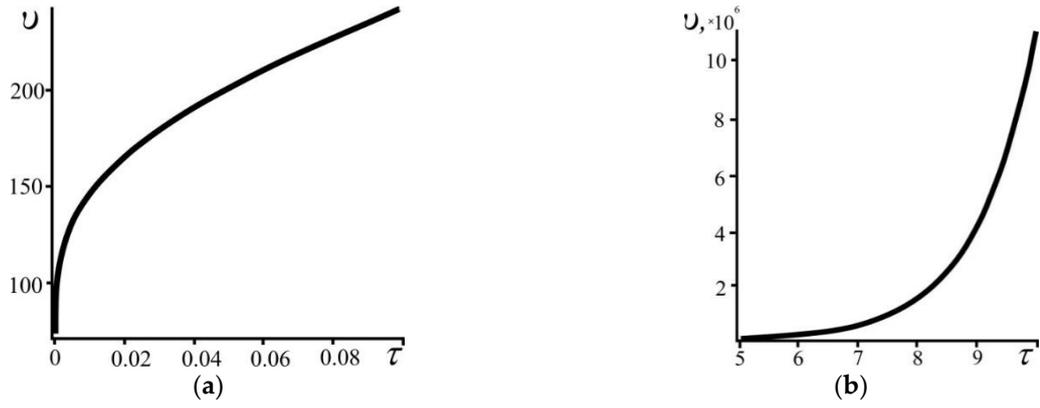

**Figure 7.** The dependence of the rate of change of the system size $v$ as a function of the time $\tau$ for the superdense (a) and the rarified (b) states. $N=10^3$.

Interestingly, for the state of low density, according to (17) and (21), the rate at which the system grows (expands) is identically equal to its size.

For the two considered particular cases, the rate of change of the system's size increases with time (Fig.7). Additionally, in the superdense state this increase occurs with decreasing acceleration, and in the state of low density with increasing acceleration.

2. Let us define the energy of the system as $E = \rho \cdot v^2$ (where $\rho=N/G$ is an average density of particles in the system). By definition, this energy is related exclusively to the rate of change of the system's size and is analogous to the kinetic energy. Based on Eq.(18), the system's energy has the general form (see Fig.8):

$$E = \frac{N^3}{G} \cdot \Omega(N,G)^{-2}. \qquad (22)$$

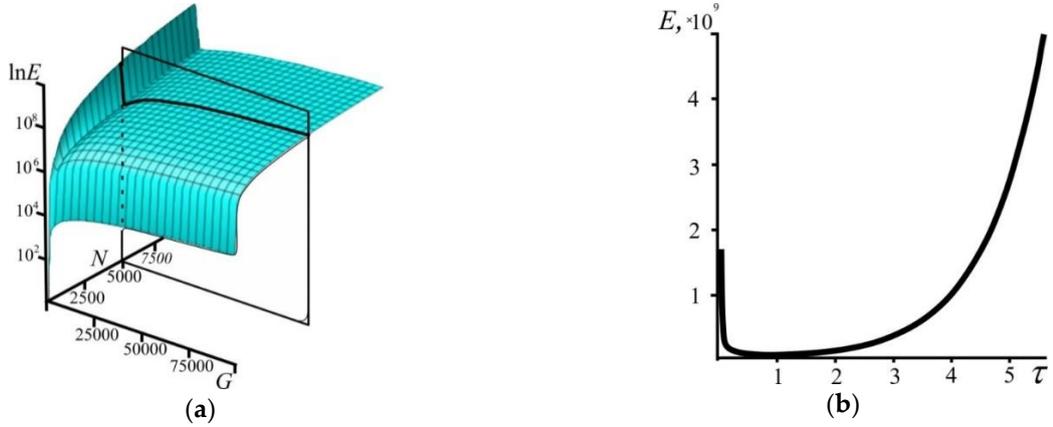

**Figure 8.** The dependence of the energy $E$ on the number of particles $N$ and the number of cells $G$ (a). For clarity, the example of the dependence $E$ on $G$ only (plane section with $N=5000$) is drawn. The dependence of the energy $E$ on the time $\tau$ (b). The dependence is plotted parametrically using Eqs.(9), (25) with $N=10^4$ and $G$ varying from 10 to $5\times10^5$.

For the two extreme cases, we obtain energy explicitly dependent on time. For the superdense state, based on (20) and (15), the system's energy is:

$$E = -\frac{N^2 \cdot W_{-1}\left(-\dfrac{\tau}{2e^{1/2}}\right)}{\tau \cdot \left(1+W_{-1}\left(-\dfrac{\tau}{2e^{1/2}}\right)\right)^2}. \qquad (23)$$

Based on (17) and (21), the system's energy for the state of low density has the form:

$$E = \frac{N^2}{2} \cdot e^t. \qquad (24)$$

As can be seen from the above, the system's energy with time reaches its minimum: it decreases in a dense state but grows exponentially during further evolution and transition to a state of low density. This minimum has no special properties (not an attractor) because we consider an evolving system where time cannot "stop" according to the introduced postulate: it always increases.

3. As mentioned above, the system's kinetic energy changes, and therefore some forces do work on the system. Let us define force as $F=-(\partial E/\partial G)_N$, then, using Eq.(22), we obtain:

$$F = \frac{N^3}{G^2} \cdot \Omega(N,G)^{-2} - \frac{2 \cdot N^4}{G^2 \cdot \sqrt{N^2+G^2}} \cdot \Omega(N,G)^{-3}. \tag{25}$$

As can be seen, the force has two parts, one of which is positive and the other negative (it can be easily proved that $\Omega$ is always positive at any $N$ and $G$). By analogy with the classical representations [1], let us consider that the first term $F_1$ in Eq.(25) corresponds to "attraction":

$$F_1 = \frac{N^3}{G^2} \cdot \Omega(N,G)^{-2}, \tag{26}$$

and the second term $F_2$ represents "repulsion" of particles in the system:

$$F_2 = -\frac{2 \cdot N^4}{G^2 \cdot \sqrt{N^2+G^2}} \cdot \Omega(N,G)^{-3}. \tag{27}$$

It is seen from Fig.9 that the attraction between particles decreases and the repulsion increases as the system grows in size. Additionally, at larger sizes both contributions reach constant values. Let us find from (26), (27) an asymptotic behavior of the forces. For the superdense state ($N/G \to \infty$ or $N>>G>>1$), the attractive and the repulsive forces are inversely proportional to the square of the system size. They are:

$$F_1 \approx \frac{N^3}{(\ln 2N)^2} \frac{1}{G^2}, \quad F_2 \approx -\frac{2N^3}{(\ln 2N)^3} \frac{1}{G^2}. \tag{28}$$

We note that at a sufficiently large $N$ the repulsive force can be neglected: $F_1>>F_2$.

For the state of low density ($N/G \to 0$ or $G>>N>>1$), the forces of attraction and repulsion are comparable:

$$F_2 = -2F_1 = -2N, \quad F = -N. \tag{29}$$

Obviously, the forces arising in the considered model have a purely kinematic origin, i.e. the forces are "born" when the measure of time is introduced and then, on this basis, particles in the growing volume of space are considered.

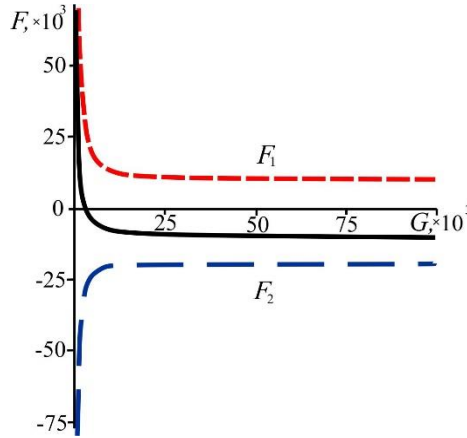

**Figure 9.** The dependence of the force of attraction $F_1$ (red dotted line), the force of repulsion $F_2$ (blue dotted line), and the total force $F$ (black solid line) on $G$ (with $N=10^4$).

## 5. Conclusion

Thus, based on the introduced entropic measure of time and using the simple model, a rather informative mechanical theory is deductively obtained with a number of interesting results, specifically: 1) a system consisting of particles of only one type evolves into a system with two types of particles having equal fractions; 2) a kinematic relationship between the size, time and number of particles of the system results in accelerated increase of the system's size and, if the system's size remains constant, in decrease of the number of particles; and 3) forces of attraction and repulsion arise which are inversely proportional to the square of the system's size for relatively dense systems and have constant values for systems with sufficiently low density.[4] Obviously, these results can be qualitatively compared with experiments, particularly in the field of astrophysics. This could be the subject of future work.

Further improvement and development of the model (with the time measure and its postulates remaining unchanged) should allow describing new observations. Potentially, the model can be developed by adding properties to the particles $b$ and $f$, e.g. charges and abilities to interact with each other. Such an iterative procedure of refining the model and comparing it with experiments could continue until a required qualitative and quantitative agreement is reached.

**Author Contributions:** Leonid M. Martyushev proposed all basic ideas of the research. He also wrote the basic text of the paper. Evgenii V. Shaiapin carried out all the necessary calculations, figures and prepared the manuscript for publication. Both authors analyzed and discussed the results, read and approved the final manuscript.

**Acknowledgments:** The work was supported by Act 211 Government of the Russian Federation, contract № 02.A03.21.0006. Part of the study was performed within scientific project No. 1.4539.2017/8.9.

**Conflicts of Interest:** The authors declare no conflict of interest.

---

[4] As was mentioned, forces arising in the present model have a purely entropic origin. It is interesting to note that in a number of fields of the modern cosmology the gravitational interaction is also related to entropy (the so-called entropic force etc.) [10-12].